\documentclass{emulateapj}
\shorttitle{21 cm-NIRB cross power spectrum during reionization}
\shortauthors{Mao}

\begin{document}

\title{Probing reionization with the cross power spectrum of 21 cm and near-infrared radiation backgrounds}

\author{Xiao-Chun Mao\altaffilmark{1} \altaffilmark{\dag}}

\altaffiltext{1}{National Astronomical Observatories, Chinese Academy of
Sciences, Beijing 100012, China}
\altaffiltext{\dag}{E-mail: xcmao@bao.ac.cn}

\slugcomment{The Astrophysical Journal, 790:148 (11pp), 2014 August 1}
\received{2013 November 7}
\accepted{2014 June 16}
\published{2014 July 16}

\begin{abstract}
The cross-correlation between the 21 cm emission from the high-redshift intergalactic medium and the near-infrared (NIR) background light from the high-redshift galaxies promises to be a powerful probe of cosmic reionization. In this paper, we investigate the cross power spectrum during the epoch of reionization. We employ an improved halo approach to derive the distribution of the density field and consider two stellar populations in the star formation model: metal-free stars and metal-poor stars. The reionization history is further generated to be consistent with the electron-scattering optical depth from cosmic microwave background measurements. Then the intensity of NIR background is estimated by collecting emission from stars in the first-light galaxies. On large scales, we find the 21 cm and NIR radiation backgrounds are positively correlated during the very early stages of reionization. However, these two radiation backgrounds quickly become anti-correlated as reionization proceeds. The maximum absolute value of the cross power spectrum is $|\Delta^2_{\rm{21,NIR}}|\sim10^{-4}$ mK nW $\rm{m}^{-2}$ $\rm{sr}^{-1}$ reached at $\ell\sim1000$, when the mean fraction of ionized hydrogen is $\bar{x}_{i}\sim0.9$. We find that SKA can measure the 21 cm-NIR cross power spectrum in conjunction with mild extensions to the existing CIBER survey, provided that the integration time independently adds up to 1000 and 1 hours for 21 cm and NIR observations, and that the sky coverage fraction of CIBER survey is extended from $4\times10^{-4}$ to 0.1. Measuring the cross-correlation signal as a function of redshift provides valuable information on reionization and helps confirm the origin of the ``missing'' NIR background.
\end{abstract}

\keywords{cosmology: theory --- large-scale structure of universe --- intergalactic medium --- diffuse radiation}

\section{Introduction}

The epoch of reionization (EoR), during which the cosmic gas went from a neutral to an ionized state, is related to many fundamental questions in cosmology, such as the formation and properties of the first stars, the escape of ionizing photons from star-forming regions, and the emergence of the Cosmic Web. Yet it still remains poorly understood. Currently, this situation is changing through several planned and ongoing experiments.

One of the most promising techniques for studying the EoR is the 21 cm hyperfine transition of neutral hydrogen. During the EoR, the first collapsed objects heated and ionized the intergalactic medium (IGM). As a direct probe of the IGM, the 21 cm line is sensitive to the emergence of the first stellar populations, the radiation from growing massive black holes and the formation of larger groups of galaxies and bright quasars. Simulations of the IGM evolution have shown that the redshifted 21 cm signal has a strength of $\sim10$ mK \citep[e.g.][]{Matteo02,Ciardi03,McQuinn06,Jelic08}. Low-frequency radio experiments such as the Low Frequency Array (LOFAR) \footnote{http://www.lofar.org}, Murchison Widefield Array (MWA) \footnote{http://www.mwatelescope.org}, 21 CentiMeter Array (21CMA) \footnote{http://21cma.bao.ac.cn}, Giant Meterwave Radio Telescope (GMRT) \footnote{http://gmrt.ncra.tifr.res.in}, Precision Array to Probe Epoch of Reionization (PAPER) \footnote{http://astro.berkeley.edu/~dbacker/eor}, and Square Kilometer Array (SKA) \footnote{http://www.skatelescope.org} will aim to seek detections of the 21 cm signal from the EoR in the near future. Unfortunately, the redshifted 21 cm line is swamped by a long list of contaminants. In addition to the experimental difficulties at low radio frequencies, the presence of Galactic and extragalactic foreground sources, which contribute a brightness temperature on the order of $\sim100$ K at 100 MHz, does pose a serious challenge for the real-world observations \citep[e.g.][]{Shaver99,Furlanetto06,Bowman09,Liu09a}. In the last decade, a great deal of attention - both observational and theoretical - has been focused on understanding the data components in these low-frequency experiments, in order to prepare us for the upcoming detections \citep[e.g.][]{Bharadwaj01,Morales08,Harker10,Liu10,Ord10,Jacobs11,Mao11,Trott12}.

Another way to observe the EoR is through the cosmic near-infrared background (NIRB) \citep{Santos02,Kashlinsky02,Cooray04,Cooray04b,Kashlinsky04,Madau05,Kashlinsky07,
Fernandez10,Kashlinsky12,Fernandez12,Fernandez13}. The first generations of stars and galaxies can produce ultraviolet (UV) radiation through nuclear reactions. These UV photons ($\lambda \sim 1000$ \AA) at redshifts $6 \lesssim z \lesssim 30$ will be redshifted into the NIR band. Such radiation is not expected to be present in the background light at UV and optical bands due to the redshifted Lyman limit. Therefore, it makes sense to search for this remnant light in the NIR band to study high-redshift galaxies and reionization. Theoretical proposals have motivated experimental measurements both on the integrated intensity and on the anisotropy power spectrum of the NIRB. After subtracting foregrounds and point sources, the sky indeed shows a faint excess emission of extragalactic origin \citep{Dwek98,Wright00,Kashlinsky05,Matsumoto05,Kashlinsky07,Matsumoto11,Kashlinsky12}. Where does the NIRB excess come from? Based on current measurements, \citet{Helgason12} suggested that this non-zero excess is unable to be attributed to the known galaxy populations. A compelling interpretation is that it originates from the many small and faint galaxies at $z>6$ \citep{Santos02,Salvaterra03,Kashlinsky05,Kashlinsky07,Matsumoto11,Kashlinsky12}. If true, the NIRB would be a powerful probe of the EoR since it would probe all sources rather than only the brightest ones. While there exists debates on the origin of the excess of the cosmic NIRB \citep{Thompson07a,Cooray07,Thompson07b,Chary08,Cooray12}, one still expects that some contribution to it is from high-redshift sources.

Continuing efforts are expended to measure the cosmic 21 cm and NIR radiation backgrounds. It is natural to consider how these two probes can be combined to reveal even more information about the reionization process. The existence of the 21 cm-NIRB cross- correlation comes from the fact that inhomogeneities that lead to spatial fluctuations in the 21 cm and NIR backgrounds trace the same underlying density field. The potential synergy of the cross-correlation study is obvious. Firstly, the 21 cm cross-correlation with NIRB will provide a more direct tracer of the interplay between the reionizing sources and the surrounding IGM, than their respective auto-correlations. During the EoR, the temperature of the IGM would be changed dramatically by photo-ionization, which had a significant impact on further galaxy formation and growth. Since the reionization process depends on the mass and clustering of ionizing sources, measuring the 21 cm-NIRB correlation throughout reionization can teach us a great deal about the nature of the first galaxy population. Beyond the reionization history, cross-correlation studies can be used to establish the presence of the cosmic NIRB radiation from high-redshift sources. Since the NIRB at a given frequency consists of the cumulative radiation from all extragalactic sources, the NIRB maps do not carry precise redshift information about radiation sources and are not expected to correlate perfectly with each other. Differently, the 21 cm emission comes from a single spectral line, indicating that the associated fluctuations can be well-localized in redshift space. Cross-correlating NIRB fields with 21 cm measurements could give a clear signal if the unresolved NIRB is dominated by the first galaxies, thus allowing us to make statements about the origin of the detected NIRB excess. Cross-correlation also has some useful properties on the data analysis side: it suffers from foregrounds and systematic effects less than auto-correlations.

In this work, we consider such a combined study. A rough estimate has been presented by \citet{Slosar07}. They used a simple analytic model to predict the qualitative features of the cross power spectrum. With the exciting progress made in the study of reionization, detailed theoretical modeling is now required to forecast the cross-correlation signal, and eventually interpret the results of future observations. Here we introduce a physically motived analytical approach to make firmer statements on how strong the correlation signal could be expected. We employ a halo model to calculate the non-linear clustering of the density field and make use of the halo occupation distribution (HOD) for the first galaxies.  Based on the ongoing star formation model, we further examine energy spectra of various emission processes from early Pop II and Pop III stars. In order to check the consistence of our stellar model and reionization history, we specially estimate the electron-scattering optical depth and compare it to the WMAP measurements \citep{Komatsu11}. Moreover, the cross-correlation between neutral hydrogen fraction and galaxy density field is taken into account in our calculations, which is significant but neglected in \citet{Slosar07}. The theoretical investigations offer the best hope of exploring the astrophysical contents of the cross-correlation, although complete simulations will be necessary to understand the entire process more properly.

The paper is organized as follows. In Section 2, we revisit the angular power spectra of cosmic 21 cm and NIR radiation backgrounds and describe the cross-correlation between them. Next, we examine the extent to which the 21 cm-NIRB correlation can be detected in Section 3. Finally, we conclude with a discussion of our results in Section 4. Throughout the paper we adopt a concordance cosmology of $\Omega_m=0.27$, $\Omega_\Lambda=0.73$, $\Omega_b=0.046$, $h=0.71$, $n_s=0.96$ and $\sigma_8=0.81$, as revealed by the WMAP seven-year observations \citep{Komatsu11}

\section{Angular power spectrum}

In this paper we are interested in measuring the cross-correlation between the cosmic 21 cm and NIR radiation backgrounds. For this purpose, here we will describe the power spectra of  21 cm fluctuations and NIRB anisotropies, and further derive the cross-correlation between them. Since the auto-correlations have been obtained by previous literature \citep[cf.][]{Zaldarriaga04,Santos05,Fernandez10,Cooray12}, we give a general derivation here and more details are available in original papers. In order to evaluate auto- and cross-correlations numerically under the analytic model, we consider scales much larger than a typical bubble size and focus on the large-scale behavior of power spectra. The flat-sky coordinates are used throughout the paper.
 
\subsection{21 cm power spectrum}

Ignoring peculiar velocities, the 21 cm brightness temperature, relative to the CMB, at observed frequency $\nu$ (i.e. redshift z) can be written as \citep[e.g.][]{Field58,Field59,Zaldarriaga04}:
\begin{eqnarray}
T_{21}(\hat{\mathbf{n}},z)&\approx& 26 \,\bar{x}_{\rm {HI}}(1+\delta_{\rm \rho})(1+\delta_{\rm x}) \Bigg(\frac{T_s-T_{\rm CMB}}{T_s}\Bigg) \Bigg(\frac{\Omega_{\rm b}h^2}{0.022}\Bigg)\nonumber \\ 
&&\times \Bigg(\frac{1+z}{10}\,\frac{0.15}{\Omega_{\rm M}h^2}\Bigg)^{1/2}\, {\rm mK}.
\end{eqnarray}
In this equation, $\bar{x}_{\rm {HI}}$ is the volume-averaged neutral hydrogen fraction. The field $\delta_{\rm \rho}$ is the fractional gas density fluctuation, while $\delta_{\rm x}=(x_{\rm {HI}}-\bar{x}_{\rm {HI}})/\bar{x}_{\rm {HI}}$ is the perturbation in the neutral hydrogen fraction. $T_s$ is the spin temperature of the 21 cm transition, and $T_{\rm CMB}$ is the CMB temperature. The other symbols have their usual meanings. Here we adopt the simplifying assumption that $T_s\gg T_{\rm CMB}$ globally, which should be a good approximation during most of the reionization epoch \citep{Mesinger07}. We can now write the observed 21 cm brightness temperature on the sky as
\begin{eqnarray}
T_{21}(\hat{\mathbf{n}},z)=T_0(z)\,\bar{x}_{\rm {HI}}\,\int dr\,W(r)\,\psi(\hat{\mathbf{n}},r),
\end{eqnarray}
where we define
\begin{eqnarray}
T_0=26\,\Bigg(\frac{\Omega_{\rm b}h^2}{0.022}\Bigg) \Bigg(\frac{1+z}{10}\,\frac{0.15}{\Omega_{\rm M}h^2}\Bigg)^{1/2}\, {\rm mK}
\end{eqnarray}
is the mean 21 cm brightness temperature of the IGM, $r$ is the radial distance around the target redshift $z$, and $\psi(\hat{\mathbf{n}},r)=(1+\delta_{\rm \rho})(1+\delta_{\rm x})$. Note that $W(r)$ is the window function representing the frequency resolution of telescope.

We first decompose the 21 cm temperature field into spherical harmonic multiple moments
\begin{eqnarray}
a^{21}_{\ell m}&=&\int d \hat{\mathbf{n}}\,T_{21}(\hat{\mathbf{n}},z)Y^{\ast}_{\ell m}(\hat{\mathbf{n}}) \nonumber\\
&=&4\pi i^{\ell}\int \frac{d^3 k}{(2\pi)^3}\,\psi_{k}(\mathbf{k})\,I^{21}_{\ell}(k,z)\,Y^{\ast}_{\ell m}(\mathbf{k})
\end{eqnarray}
in which
\begin{eqnarray}
I^{21}_{\ell}(k,z)=\int dr\,W^{21}(r,z)j_{\ell}(kr),
\end{eqnarray}
with the weighting function
\begin{eqnarray}
W^{21}(r,z)=T_0(z)\,\bar{x}_{\rm {HI}}\,W(r).
\end{eqnarray}

Next we calculate the angular power spectrum of 21 cm fluctuations, which is defined by the two-point correlation function
\begin{eqnarray}
\langle a^{21}_{\ell m}a^{21}_{\ell' m'}\rangle=\delta_{\ell \ell'}\delta_{mm'}C^{\psi\psi}_{\ell}.
\end{eqnarray}
Using the weighting function in radial space, we can finally construct the angular power spectrum as a line of sight projection \citep{Song03}
\begin{eqnarray}
C^{\psi\psi}_{\ell}(z)=\int \frac{dr}{r^2}\Big[W^{21}(r,z)\Big]^2\,P_{21}(k=\frac{\ell}{r},z).
\end{eqnarray}
We have used the Limber approximation \citep{Limber54} here. Furthermore, the 21 cm power spectrum $P_{21}(k,z)$ can be decomposed into the sum of several terms (generalizing the formula in \citet{Furlanetto06b})
\begin{eqnarray}
P_{21}(k,z) &=& P_{\rm{\rho,\rho}}(k,z)+P_{\rm{x,x}}(k,z)+2P_{\rm{\rho,x}}(k,z)+2P_{\rm{x\rho,x}}(k,z) \nonumber \\
&& +2P_{\rm{x\rho,\rho}}(k,z)+P_{\rm{x\rho,x\rho}}(k,z).
\end{eqnarray}
The quantity $P_{\rm{a,b}}$ indicates the dimensional power spectrum of two random fields, $a$ and $b$. As shown in \citet{Lidz06}, the last three terms, referred to as the higher-order terms, contribute significantly to the 21 cm power spectrum only at small scales $k>1\,{\rm Mpc}^{-1}$ and hence are ignored in our calculations. The first three terms represent the matter power spectrum, the power spectrum of neutral hydrogen fluctuations and the cross-correlation power, respectively. We have neglected the cross-correlation $P_{\rm{\rho,x}}(k,z)$ because it is small in the scales of interest \citep{Santos03}.

Under the halo model, we can write the matter power spectrum as \citep{Seljak00,Cooray02}
\begin{eqnarray}
P_{\rm {\rho,\rho}}(k,z)&=&P^{1h}_{\rm {\rho,\rho}}(k,z)+P^{2h}_{\rm {\rho,\rho}}(k,z) \nonumber \\
P^{1h}_{\rm {\rho,\rho}}(k,z)&=&\int dM \frac{dn(z,M)}{dM}\,\Bigg|\frac{\rho_h(z,M,k)}{\bar{\rho}(z)}\Bigg|^2 \nonumber \\
P^{2h}_{\rm {\rho,\rho}}(k,z)&=&\int dM_1\,\frac{dn(z,M_1)}{dM_1}\,\frac{\rho_h(z,M_1,k)}{\bar{\rho}(z)}\,b_h(z,M_1) \nonumber \\
&&\times \int dM_2\,\frac{dn(z,M_2)}{dM_2}\,\frac{\rho_h(z,M_2,k)}{\bar{\rho}(z)}\,b_h(z,M_2) \nonumber \\
&&\times P_{\rm {lin}}(k,z).
\end{eqnarray}
Here $M$ is the halo mass, $dn/dM(z,M)$ is the halo mass function \citep{Sheth99}, $\rho_h(z,M,k)$ is the Fourier transform of the NFW halo density profile \citep{Navarro97}, $\bar{\rho}(z)$ is the cosmic mean matter density, $b_h(z,M_)$ is the halo bias and $P_{\rm {lin}}(k,z)$ denotes the linear matter power spectrum. For precise calculations, we have included the baryon content in the computation of matter power spectrum.

Considering the case where $\bar{x}_{\rm {HI}}\,\delta_{\rm x}=-\bar{x}_{\rm i}\,\delta_{\rm i}$, $P_{\rm{x,x}}$ can be expressed in terms of the ionization fraction equation:
\begin{eqnarray}
{\bar{x}_{\rm {HI}}}^2\,P_{\rm{x,x}}(k,z)={\bar{x}_{\rm i}}^2\,P_{\rm{i,i}}(k,z).
\end{eqnarray}
Here $\delta_{\rm i}=(x_{\rm i}-\bar{x}_{\rm i})/\bar{x}_{\rm i}$ is the perturbation in the ionization fraction and $P_{\rm{i,i}}(k,z)$ is the power spectrum from $\delta_{\rm i}$. Under the assumption that ionizing radiation was coming from stars formed from gas clouds that cooled in dark matter halos, we are thus able to write the auto correlation of ionization fraction field as a multiple of the dark matter correlation \citep{Santos03,Santos05}
\begin{eqnarray}
P_{\rm{i,i}}(k,z)={b^2_{\rm {eff}}}(z)\,P_{\rm{\rho,\rho}}(k,z)\,e^{-k^2R^2},
\end{eqnarray}
where $b_{\rm {eff}}(z)$ is the mean bias weighted by the different halo properties and $R$ is the mean radius of the HII patches. We follow the same approach as \citet{Santos03} and \citet{Santos05} for these two variables.

We now need to calculate the hydrogen ionization fraction $\bar{x}_{\rm i}$ as a function of redshift. Following \citet{Cooray12}, $\bar{x}_{\rm i}$ can be estimated as
\begin{eqnarray}
\frac{d \bar{x}_{\rm i}}{d t}=\frac{f_{\rm{esc}}\,\varphi(z)q(z)}{\bar{n}_{\rm H}(z)}-\frac{\bar{x}_{\rm i}}{\bar{t}_{\rm{rec}}}
\end{eqnarray}
in which $f_{\rm{esc}}$ is the escape fraction of ionization photons and $\bar{n}_{\rm H}(z)=1.905\times10^{-7}(1+z)^3\,\rm{cm}^{-3}$ is the mean hydrogen number density \citep{Shull11}. Here, $\varphi(z)$ is the comoving star formation rate density (SFRD) given by the ongoing star formation model \citep{Santos02}. The function $q(z)$ is defined as $q(z)\equiv(\bar{Q}_{\rm{HI}}/\langle M_{\ast} \rangle)\,\langle \tau_{\ast} \rangle$, where $\bar{Q}_{\rm{HI}}$ is the time-averaged hydrogen photoionization rate, $\langle M_{\ast} \rangle$ is the average star mass and $\langle \tau_{\ast} \rangle$ is the average stellar lifetime \citep{Lejeune01,Schaerer02}. The $\bar{t}_{\rm{rec}}$ denotes the volume averaged recombination time \citep{Madau99}.

Figure 1 shows the evolution of the hydrogen ionization fraction with redshift. Here and throughout we set the star formation efficiency $f_{\ast}=0.04$, the escape fraction of ionization photons $f_{\rm esc}=0.5$ and a gas temperature $T=3\times10^4\,\rm K$. In addition, $M_{\rm min}=10^6\,\rm{M}_{\odot}$ and $M_{\rm max}=10^{14}\,\rm{M}_{\odot}$ denote the minimum and maximum halo mass to host galaxies respectively. With these parameters we find that reionization ends around a redshift of 6 which is consistent with current studies. Furthermore, we follow the approach of \citet{Cooray12} to estimate the electron-scattering optical depth with the ionization fraction. We find that the derived optical depth is $\tau=0.084$ which is close to the result of WMAP seven-year data with $\tau=0.088\pm0.014$ \citep{Komatsu11}.

\begin{figure}
\begin{center}
\includegraphics[angle=270, scale=0.5]{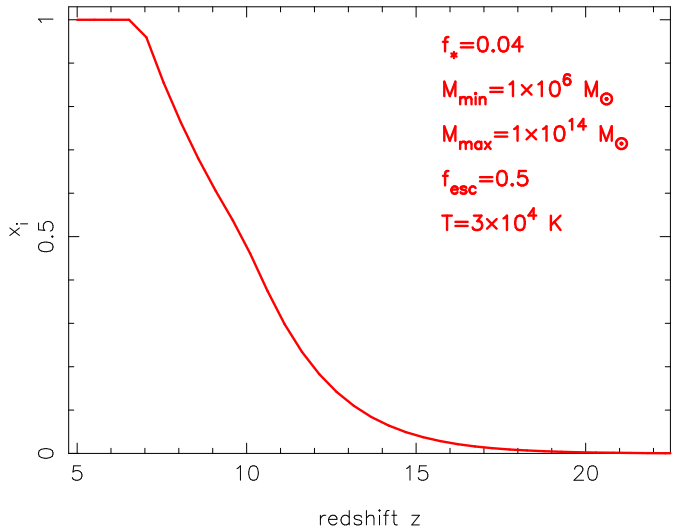}
\caption{Hydrogen ionization fraction as a function of redshift.}
\end{center}
\end{figure}

\subsection{NIRB power spectrum}

Similar to \citet{Knox01}, we write the antenna temperature of the NIRB at a given frequency $\nu$ and towards a direction $\hat{\mathbf{n}}$ as a product of the mean NIRB emissivity and its fluctuation
\begin{eqnarray}
T_{\rm NIR}(\hat{\mathbf{n}},\nu)=\int dr\,a(r)\,\bar{j}(\nu,r)\,\Bigg[1+\frac{\delta j(r\hat{\mathbf{n}},\nu,r)}{\bar{j}(\nu,r)}\Bigg].
\end{eqnarray}
Here $r$ is the radial distance from us to a redshift of $z$, $a=(1+z)^{-1}$ is the scale factor and $\bar{j}(\nu,z)$ is the mean emissivity per comoving unit volume at frequency $\nu$ and redshift $z$. To model the spatial fluctuations related to the emissivity, we assume that the sources of this emissivity are galaxies, such that $\delta j(r\hat{\mathbf{n}},\nu,z)/\bar{j}(\nu,z)=\delta_{\rm gal}$ where $n_{\rm gal}=\bar{n}_{\rm gal}(1+\delta_{\rm gal})$ is the comoving number density of galaxies. Note that the NIRB would probe all galaxies as a whole, not just the brightest ones which can be detected individually in galaxy surveys. For a specific NIR experiment, we integrate $\bar{j}(\nu,r)$ over the observed bandpass and further define the observed NIRB temperature as
\begin{eqnarray}
T_{\rm NIR}(\hat{\mathbf{n}})=\int dr\,a^2(r)\,\bar{j}(r)\,\phi(\hat{\mathbf{n}},r),
\end{eqnarray}
in which $\bar{j}(z)$ is the band-averaged mean emissivity and $\phi(\hat{\mathbf{n}},r)=1+\delta_{\rm gal}$. Note that here we have a factor $a^2$ instead of $a$ in Equation (14) since we are no longer estimating the temperature at only one frequency. This dependence has been explained in the Appendix of \citet{Fernandez10}.

In this work, we consider the emission from stars in first-light galaxies present during reionization and neglect the emission from the IGM. There are two stellar populations in our calculation. The first, referred to as Pop II stars, are metal-poor stars with metallicity $Z=1/50\,Z_{\odot}$, and the second, Pop III stars, are metal-free stars with $Z=0$. For Pop II stars, we adopt the stellar initial mass functions (IMF) given by \citet{Salpeter95} with the mass range from $3$ to $150\,M_{\odot}$. For Pop III stars, we make use of the IMF obtained by \citet{Larson99}, and the mass range is from $3$ to $500\,M_{\odot}$. Collecting all the emission from these two stellar populations, we obtain the band-averaged mean emissivity as
\begin{eqnarray}
\bar{j}(z)=f_{\rm P}\,\bar{j}^{\rm {PopIII}}(z)+(1-f_{\rm P})\,\bar{j}^{\rm{PopII}}(z).
\end{eqnarray}
Here $f_{\rm P}(z)$ is the population fraction describing the relative fraction of the Pop II and Pop III stars at different redshifts \citep{Cooray12}. The functions $\bar{j}^{\rm {PopII}}(z)$ and $\bar{j}^{\rm{PopIII}}(z)$ denote the band-averaged mean emissivities for Pop II and Pop III stars respectively \citep{Cooray12}
\begin{eqnarray}
\bar{j}^{i}(z)=\frac{1}{4\pi}\,\bar{l}^{i}\,\langle\tau^{i}_{\ast}\rangle\,\varphi(z),
\end{eqnarray}
where $\langle\tau_{\ast}\rangle$ is the mean stellar lifetime of each of the stellar type and $\varphi(z)$ is the comoving star formation rate density (SFRD). We integrate the luminosity mass density $l_{\nu}$ over the band of observed frequencies $\nu_1$ to $\nu_2$ to obtain the band-averaged $\bar{l}$ as \citep{Fernandez10}


    \begin{deluxetable}{p{1.8cm}p{1.8cm}p{1.8cm}p{1.8cm}}
    \tablewidth{18pc}
    \tablecaption{Band definitions used for NIR observations}
   \tablehead{Band & $\lambda_1\,(\mu m)$ & $\lambda_2\,(\mu m)$ & $\lambda_{\rm{obs}}\,(\mu m)$}

    \startdata
    H & 1.5 & 1.8 & 1.65 \\
    L & 3.0 & 4.0 & 3.5 \\
    \enddata

    \tablecomments{$(\lambda_1$,$\lambda_2)$ defines the waveband in NIR observation, corresponding to the observed bandpass $(\nu_2$,$\nu_1)$. $\lambda_{\rm{obs}}$ is a convenient quantity that tells us what the center wavelength of the observed waveband is.}
    \end{deluxetable}

\begin{figure*}
\begin{center}
\includegraphics[angle=270, scale=0.7]{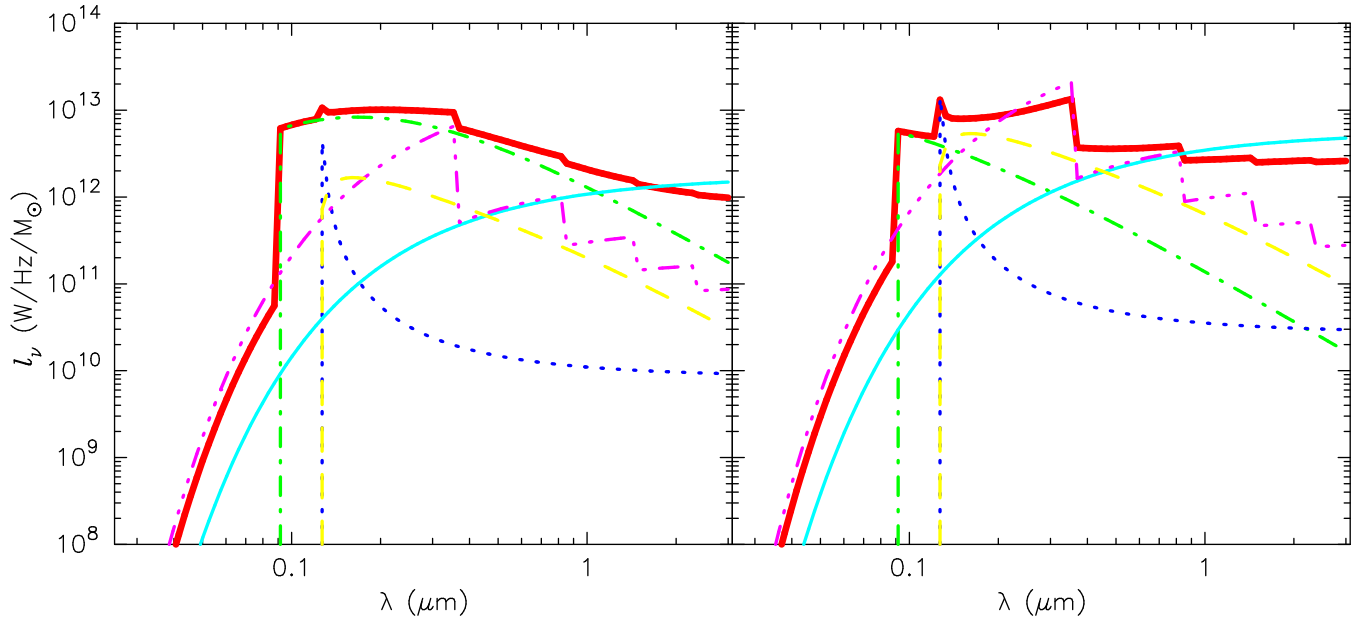}
\caption{Luminosity mass density for Pop II (left) and Pop III (right) stars at $z=10$. The stellar (dot dashed line), Lyman-$\alpha$ (dotted line), free-free (thin solid line), free-bound (triple-dot dashed line) and two photon (dashed line) emission are shown as a function of the rest-frame wavelength $\lambda$, respectively. The total $l_{\nu}$ from all these sources is also plotted as the thick solid line in each panel. Parameters are the same as those adopted in Figure 1.}
\end{center}
\end{figure*}

\begin{eqnarray}
\bar{l}(z)=\int_{\nu_1(1+z)}^{\nu_2(1+z)} d\nu\,l_{\nu}(z).
\end{eqnarray}
Following previous literature \citep{Fernandez06,Fernandez10,Cooray12}, we examine the luminosity mass density $l_{\nu}(z)$ for various emission processes that contribute to the NIRB, such as stellar emission, Lyman-$\alpha$ line and free-free, free-bound and two-photon emission. Here we do not present the detailed theoretical calculations of the luminosity mass density for each radiative process and the basic formulae can be found in \citet{Fernandez06}. We finally derive the total luminosity mass density from the stellar nebulae
\begin{eqnarray}
l_{\nu}=l^{\ast}_{\nu}+(1-f_{\rm{esc}})\big(l^{\rm{Ly\alpha}}_{\nu}+l^{\rm{ff}}_{\nu}+l^{\rm{fb}}_{\nu}+l^{\rm{2ph}}_{\nu}\big),
\end{eqnarray}
where $f_{\rm{esc}}=0.5$ is the escape fraction of ionization photons and $l^{i}_{\nu}$ denotes the luminosity mass density for a certain emission process. The relevant stellar parameters, such as the intrinsic bolometric luminosity, the effective temperature, the main-sequence lifetime and the time-averaged hydrogen photoionization rate, are calculated using the fitting results from \citet{Lejeune01} and \citet{Schaerer02}. Moreover, the NIR bandpasses $\nu_1$ to $\nu_2$, which are taken from \citet{Sterken92}, are defined in Table 1.

In Figure 2, we show the luminosity mass density $l_{\nu}$ as a function of the rest-frame wavelength $\lambda$ for Pop II and Pop III stars at $z=10$. To specify the various emission processes contributing to the NIRB, we plot the $l_{\nu}$ of stellar emission (dot dashed line), Lyman-$\alpha$ line (dotted line), and free-free (thin solid line), free-bound (triple-dot dashed line) and two photon (dashed line) emission. For Pop II stars (left panel), we can see that the stellar emission is dominant over the wavelength range $\lambda \sim 0.1-1\,\mu m$. However, the reprocessed light, such as free-bound and two photon, is comparable with or even larger than the stellar spectrum for Pop III stars (right panel). As illustrated in these two panels, the total $l_{\nu}$ (thick solid line) from Pop II stars is similar to that from Pop III stars. Similar results were measured in \citet{Fernandez06}.

If we expand $\phi(\hat{\mathbf{n}},r)$ in Fourier series and use the spherical harmonic decomposition in exact analogy with Equations (4)-(8), we can define the NIRB angular power spectrum given by
\begin{eqnarray}
C^{\phi\phi}_{\ell}&=&\int \frac{dr}{r^2}\Big[W^{\rm NIR}(z)\Big]^2\,P_{\rm {gal,gal}}(k=\frac{\ell}{r},z) \nonumber \\
&=&\int_{z_{\rm {min}}}^{z_{\rm {max}}} dz\,\frac{dr}{dz}\,\frac{1}{r^2}\Big[W^{\rm {NIR}}(z)\Big]^2 P_{\rm {gal,gal}}(k=\frac{\ell}{r},z).
\end{eqnarray}
Here
\begin{eqnarray}
W^{\rm NIR}(z)=a^2(z)\,\bar{j}(z)
\end{eqnarray}
is the weighting function. Here, we examine the high-redshift component of the NIRB, coming from the first galaxies during reionization. For NIR observation at H band, we calculate the angular power spectrum by integrating Equation (20) over the redshift range from $z_{\rm {min}}=6$ to $z_{\rm{max}}=17$. We do not consider the emission comes from sources at $z > 17$ since at the center wavelength $1.65\,\mu \rm m$, photons should be more energetic than $h\nu=13.6\,\rm{eV}$ in the rest frame beyond $z=17$. For NIR observation at L band, the redshift interval of the NIRB integral is $(z_{\rm {min}},z_{\rm{max}})=(6,30)$. The function $P_{\rm {gal,gal}}(k,z)$ is the power spectrum of first galaxies. The spatial distribution of these galaxies is described through the relation of the tracer field to the dark matter halo distribution. Besides halo model, the other important ingredient is how these galaxies occupy dark matter halos. We use the halo occupation distribution (HOD) model for the first generation of galaxies. Then the galaxy power spectrum can be written as
\begin{eqnarray}
P_{\rm {gal,gal}}(k,z)&=&P^{1h}_{\rm {gal,gal}}(k,z)+P^{2h}_{\rm {gal,gal}}(k,z).
\end{eqnarray}
Here $P^{1h}_{\rm {gal,gal}}$ and $P^{2h}_{\rm {gal,gal}}$ denote the power spectra contributed by galaxies in a single dark matter halo and galaxies in two different dark matter halos respectively, which can by given by \citep{Cooray02}
\begin{eqnarray}
P^{1h}_{\rm {gal,gal}}(k,z)&=&\int dM \frac{dn(z,M)}{dM}\,\frac{\langle N_{\rm {gal}}(M)[N_{\rm {gal}}(M)-1]\rangle}{\bar{n}^2_{\rm {gal}}(z)} \nonumber \\
&&\times \Bigg[\frac{\rho_h(z,M,k)}{M}\Bigg]^p \nonumber \\
P^{2h}_{\rm {gal,gal}}(k,z)&=&\Bigg[\int dM\,\frac{dn(z,M)}{dM}\, \frac{\langle N_{\rm {gal}}(M)\rangle}{\bar{n}_{\rm {gal}}(z)}\,\frac{\rho_h(z,M,k)}{M} \nonumber \\
&&\times\,b_h(z,M)\Bigg]^2\,P_{\rm {lin}}(k,z).
\end{eqnarray}
We make use of $N_{\rm {gal}}(M)$ to indicate the average number of galaxies in each dark matter halo of mass $M$. The $\bar{n}_{\rm {gal}}(z)$ is the mean number density of galaxies. The parameter $p=1$ when $\langle N_{\rm {gal}}(N_{\rm {gal}}-1)\rangle\leq1$ and $p=2$ otherwise \citep{Cooray02}.

In Figure 3, we show the 21 cm (top) and NIRB (bottom) anisotropy angular power spectra predicted by the theoretical model. Here $\Delta^2=\ell(\ell+1)C_{\ell}/2\pi$ denotes the dimensionless power spectrum. For 21 cm experiment, we take the width of one frequency bin $\Delta\nu=0.1\,\rm {MHz}$. Note that the amplitudes of $\Delta_{\rm NIR}$ shown here are nearly 10 times smaller than those shown in Figure 12 of \citet{Cooray12}. This is because we represent the band-averaged power spectrum rather than that at only one wavelength. As mentioned above, the difference between our Equation (15) and the calculation in \citet{Cooray12} is a factor of $a=1/(1+z)$, which is just $\sim 0.1$ during the EoR. Finally, the shot noise is ignored in our computation since it is much smaller than the expected NIRB signal at $l<10^4$ \citep{Fernandez10,Cooray12}.

\subsection{cross power spectrum}

Based on the derived auto-correlations of 21 cm and NIR backgrounds, we examine their cross-correlation following \citet{Song03}
\begin{eqnarray}
C^{\psi\phi}_{\ell}(z)=\int \frac{dr}{r^2}\,W^{21}(r,z)\,W^{\rm NIR}(z)\,P_{\rm {21,gal}}(k=\frac{\ell}{r},z).
\end{eqnarray}
In this equation $P_{\rm 21,gal}(k,z)$ denotes the cross power spectrum between the 21 cm fluctuation and the galaxy overdensity, which can be further decomposed into the sum of several contributing terms \citep{Lidz09}
\begin{eqnarray}
P_{\rm {21,gal}}(k,z)=P_{\rm {\rho,gal}}(k,z)+P_{\rm {x,gal}}(k,z)+P_{\rm {x\rho,gal}}(k,z).
\end{eqnarray}
The individual terms on the right-hand side have the following physical interpretations. The first term, $P_{\rm {\rho,gal}}(k,z)$, is the cross power spectrum between the matter and galaxy overdensity fields. The second term, $P_{\rm {x,gal}}(k,z)$, represents the cross power spectrum between the neutral hydrogen fraction and galaxy overdensity fields. The final term, $P_{\rm {x\rho,gal}}$, is a three-field term which contributes significantly only on small scales \citep{Lidz09} and hence is ignored in our calculation.

\begin{figure}
\begin{center}
\includegraphics[angle=270, scale=0.55]{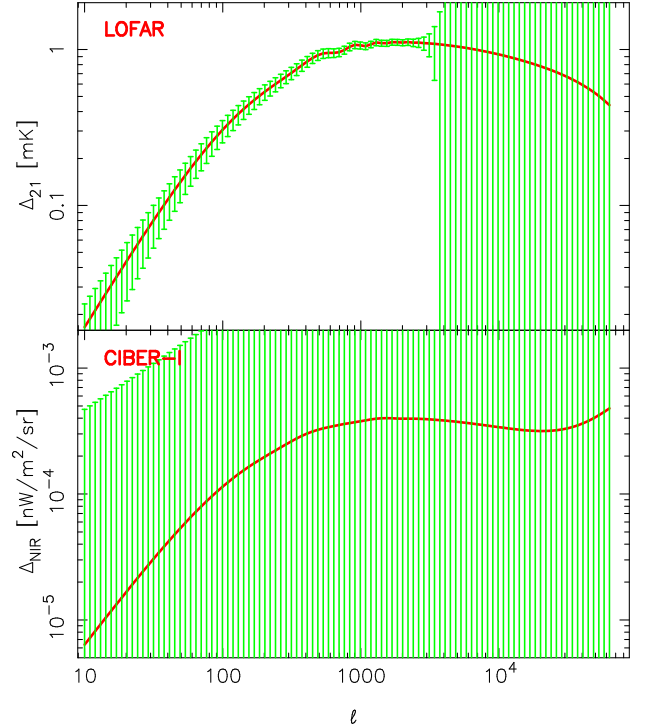}
\caption{Top panel: angular power spectrum of 21 cm fluctuations at redshift $z=10$. Errors in measurements of $\Delta_{\rm{21}}$ are estimated for an integration time of 1000 hours. Bottom panel: angular power spectrum of NIRB fluctuations at the observed H-band. Errors in measurements of $\Delta_{\rm{NIR}}$ are estimated for an integration time of 1 hour.}
\end{center}
\end{figure}

The halo model provides a simple calculation of the cross power spectrum between matter and galaxies, and we have \citep{Seljak00,Cooray02}
\begin{eqnarray}
P_{\rm {\rho,gal}}(k,z)&=&P^{1h}_{\rm {\rho,gal}}(k,z)+P^{2h}_{\rm {\rho,gal}}(k,z) \nonumber \\
P^{1h}_{\rm {\rho,gal}}(k,z)&=&\int dM \frac{dn(z,M)}{dM}\,\frac{M}{\bar{\rho}(z)}\,\frac{\langle N_{\rm {gal}}(M)\rangle}{\bar{n}_{\rm {gal}}(z)}\,\Bigg[\frac{\rho_h(z,M,k)}{M}\Bigg]^p \nonumber \\
P^{2h}_{\rm {\rho,gal}}(k,z)&=&\int dM_1\,\frac{dn(z,M_1)}{dM_1}\,\frac{\rho_h(z,M_1,k)}{\bar{\rho}(z)}\,b_h(z,M_1) \nonumber \\
&&\times \int dM_2\,\frac{dn(z,M_2)}{dM_2}\, \frac{\langle N_{\rm {gal}}(M_2)\rangle}{\bar{n}_{\rm {gal}}(z)}\,\frac{\rho_h(z,M_2,k)}{M_2} \nonumber \\
&&\times \,b_h(z,M_2)\,P_{\rm {lin}}(k,z).
\end{eqnarray}
As described above, the matter-galaxy cross-correlation has the Poisson and halo-halo terms. $P^{1h}_{\rm {\rho,gal}}(k,z)$ denotes the correlation between galaxies and dark matter in the same halo and dominates on small scales. $P^{2h}_{\rm {\rho,gal}}(k,z)$ includes the correlations between galaxies and dark matter in neighbouring halos and is dominant on large scales.

\begin{figure}
\begin{center}
\includegraphics[angle=270, scale=0.55]{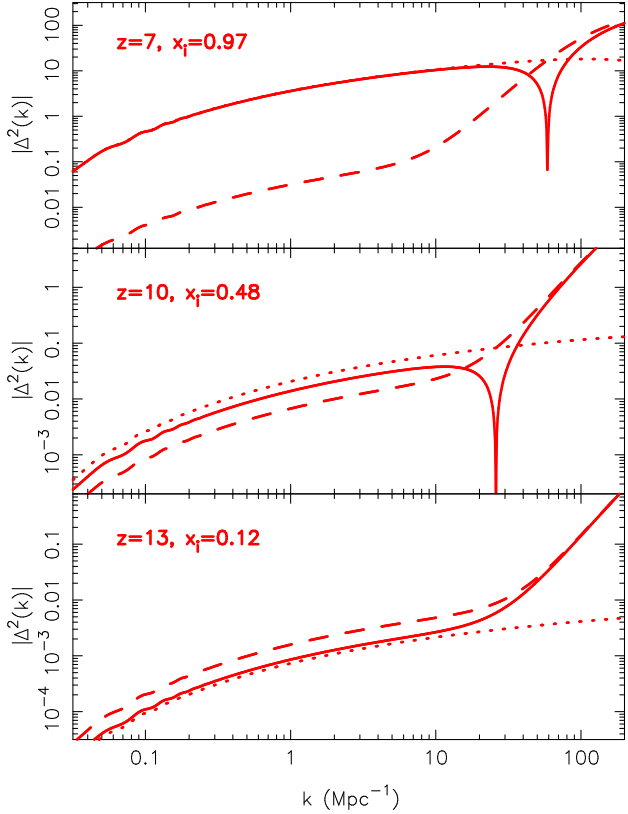}
\caption{Redshift evolution of the 21 cm-galaxy cross power spectrum $|\Delta^2_{\rm{21,gal}}|$ (solid lines). Also shown are contributions from dark matter-galaxy cross power spectrum $\Delta^2_{\rm{\rho,gal}}$ (dashed lines) and neutral fraction-galaxy cross power spectrum $|\Delta^2_{\rm{x,gal}}|$ (dotted lines). Note that we plot the absolute values of $\Delta^2_{\rm{x,gal}}$ and $\Delta^2_{\rm{21,gal}}$. While $\Delta^2_{\rm{\rho,gal}}$ is always positive, an anti-correlation is measured for $\Delta^2_{\rm{x,gal}}$ at every $k$.}
\end{center}
\end{figure}

The next step is to compute the cross-correlation between the neutral hydrogen fraction and galaxy overdensity fields. In order to evaluate $P_{\rm {x,gal}}$ numerically under the analytic model, we focus on its behavior on scales much larger than bubble sizes. With this simplification, we can rewrite Equation (12) as 
\begin{eqnarray}
P_{\rm{i,i}}(k,z)&=&b^2_{\rm {eff}}\,P_{\rho,\rho}(k,z) \nonumber \\
&=&b^2_{\rm {eff}}\,\bar{b}^2_h\,P_{\rm {lin}}(k,z),
\end{eqnarray}
where  $\bar{b}_h$ is the mean halo bias. Note that the filter function in Equation (12) is no longer included here. On large scales, the galaxy power spectrum can also be simplified to \citep{Cooray02}
\begin{eqnarray}
P_{\rm {gal,gal}}(k,z)=\bar{b}^2_{\rm {gal}}\,P_{\rm {lin}}(k,z),
\end{eqnarray}
in which $\bar{b}_{\rm {gal}}$ denotes the mean bias factor of galaxy population. Then the cross-correlation between the ionization fraction and galaxy density fields, $P_{\rm {i,gal}}$, can be given by
\begin{eqnarray}
P_{\rm {i,gal}}(k,z)=b_{\rm {eff}}\,\bar{b}_h\,\bar{b}_{\rm {gal}}\,P_{\rm {lin}}(k,z).
\end{eqnarray}
However, as described in Equation (25), $P_{\rm {21,gal}}(k,z)$ depends on $P_{\rm {x,gal}}$. This is obviously $\bar{x}_{\rm {HI}}\,P_{\rm {x,gal}}(k,z)=-\bar{x}_{\rm i}\,P_{\rm {i,gal}}(k,z)$. $\delta_x$ and $\delta_{\rm{gal}}$ are therefore anti-correlated.

Figure 4 shows the 21 cm-galaxy cross power spectrum as well as its components for $z=(7,10,13)$, and the mean fractions of ionized hydrogen are $\bar{x}_{i}=(0.97,0.48,0.12)$ correspondingly. Here the dimensionless cross power spectrum is defined as $\Delta^2_{\rm{a,b}}(k)=k^3P_{\rm{a,b}}/2\pi^2$. The dashed lines represent the density-galaxy cross power spectrum, $\Delta^2_{\rm{\rho,gal}}$. We find that this power is always positive and increased by roughly four orders of magnitude over the scale range $0.1<k<100\,\rm{Mpc}^{-1}$ at redshift $z=10$. Compared to the matter power spectrum, $\Delta^2_{\rm {\rho,\rho}}$, these high-redshift galaxies shows very strong clustering even on large scales: the cross power spectrum has an amplitude of $\Delta^2_{\rm{\rho,gal}} \simeq 6.6\times 10^{-3}$ on a scale of $k=1\,\rm{Mpc}^{-1}$ at $z=10$. On the same scale the amplitude of the matter power spectrum is $\Delta^2_{\rm {\rho,\rho}} \simeq 8.1\times 10^{-4}$, which is $\sim 8$ times smaller than that of $\Delta^2_{\rm{\rho,gal}}$. On small scales, the cross power spectrum $\Delta^2_{\rm{\rho,gal}}$ grows exponentially, indicating that structure formation is highly biased to over-dense regions during the EoR.

We plot the absolute value of the cross power spectrum between neutral hydrogen fraction and galaxy density, $|\Delta^2_{\rm{x,gal}}|$, as the dotted lines. It is worth remarking that $\Delta^2_{\rm{x,gal}}$ is negative at every value of $k$ throughout the EoR. This occurs because galaxies form first and ionize their surroundings in overdense regions, whereas underdense regions are still mostly neutral and free of galaxies. Therefore, an anti-correlation between neutral fraction and galaxy fields is naturally expected especially on large scales. One can see that the strength of $|\Delta^2_{\rm{x,gal}}|$ grows with wavenumber. Moreover, as reionization proceeds, the cross power is enhanced rapidly and finally exceeds the matter-galaxy cross power spectrum.

From Equation (25), it is clear that the $\Delta^2_{\rm{\rho,gal}}$ term and the $\Delta^2_{\rm{x,gal}}$ term will together determine the quantity of $\Delta^2_{\rm{21,gal}}$. We plot the absolute value of the 21 cm-galaxy cross power spectrum as the solid lines in Figure 4. Near the beginning of reionization, the 21 cm and galaxy density fields are positively correlated. However, these two fields quickly become anti-correlated on large scales. In our model, an interesting turnover appears when $z\sim11$ with $\bar{x}_{i}\sim0.3$ and still persists at lower redshifts. Such turnover does not exist in the 21cm-galaxy cross power spectrum at early time ($z=13,\,\bar{x}_{i}=0.12$). At $z=10$, $\Delta^2_{\rm{21,gal}}$ turns over on a scale of $k \sim 30\,\rm{Mpc}^{-1}$. The turnover behavior occurs on progressively smaller scales as reionization proceeds. Since we use an approximate expression for $\Delta^2_{\rm{x,gal}}$, the turnover presented here should be considered as a qualitative feature. Actually, the specific shape of turnover depends on the details of the reionization model \citep{Lidz09,Wiersma12}. Furthermore, one may note that the amplitudes of $\Delta^2(k)$ shown here are slightly smaller than those shown in the simulations from \citet{Lidz09}, likely owing to the smaller value of low-mass cutoff $M_{\rm min}=10^6\,\rm{M}_{\odot}$ adopted in our model of galaxy formation \citep{Scoccimarro00}.

\begin{figure*}
\begin{center}
\includegraphics[angle=270, scale=0.7]{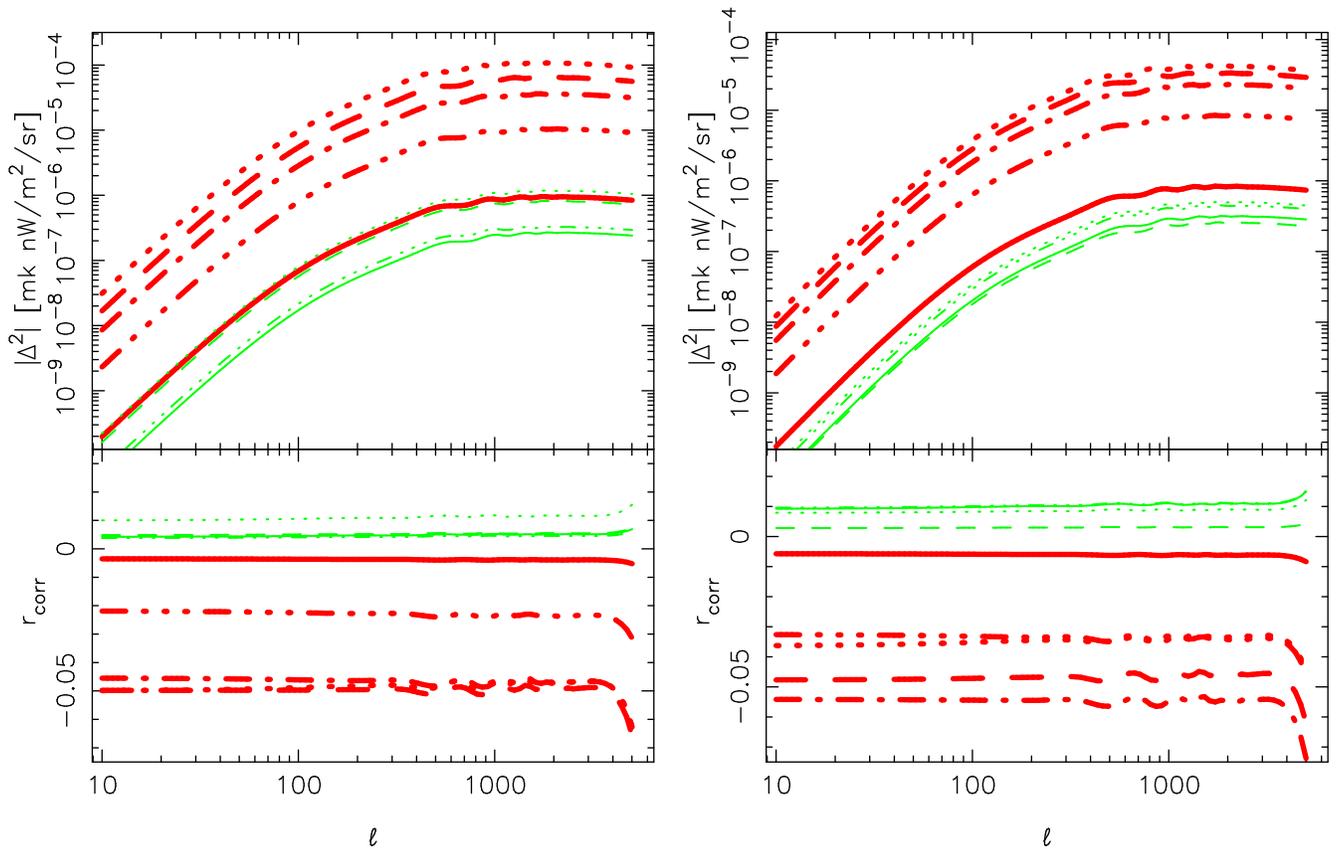}
\caption{Cross-correlation angular power spectra of the 21 cm anisotropy at individual redshifts between 7 and 15 and the cumulative NIRB during the EoR at observed H-band (left) and L-band (right). Top panels: The absolute values of the 21 cm-NIRB cross power spectra. At early stages of reionization, the 21 cm and NIR radiation backgrounds are positively correlated on large scales, as shown by the green-thin lines. At middle and late stages of reionization, these two fields are anti-correlated, as shown by the red-thick lines. Bottom panels: The cross-correlation coefficient $r_{\rm{corr}}$. In each panel, the red-thick dotted line, dashed line, dash-dot-dash-dot line, dash-dot-dot-dot line and solid line show results for $(z,\bar{x}_{i})=$ $(7,0.97)$; $(8,0.77)$; $(9,0.62)$;$(10,0.48)$; $(11,0.31)$, and the green-thin dashed line, dotted line, dash-dot-dot-dot line and solid line show results for $(z,\bar{x}_{i})=$ $(12,0.2)$; $(13,0.12)$; $(14,0.07)$; $(15,0.04)$. Note that in the bottom left-hand panel, the dashed and dash-dot-dot-dot lines happen to lie on top of each other.}
\end{center}
\end{figure*}

In addition, we use the cross-correlation coefficient to quantify the relative strength of the cross-correlation between 21 cm and NIR radiation backgrounds
\begin{eqnarray}
r_{\rm {corr}}(\ell)=\frac{C^{\psi\phi}_{\ell}}{\sqrt{C^{\psi\psi}_{\ell}\,C^{\phi\phi}_{\ell}}}.
\end{eqnarray}
In figure 5, we plot predictions for the 21 cm-NIRB cross power spectrum as well as the cross-correlation coefficient. We correlate the high-redshift component of the NIRB observed at H-band (left) and L-band(right) with the 21 cm anisotropy at individual redshifts $z=(7,8,9,10,11,12,13,14,15)$ respectively. The absolute values of cross-correlation signals are shown in upper panels, where we have $\Delta^2=\ell(\ell+1)C^{\psi\phi}_{\ell}/2\pi$. In our model, the 21 cm and NIR backgrounds are positively correlated in the early stages of reionization. At early time, galaxies are extremely rare objects and are only just starting to ionize their surroundings. The galaxies turn on first in large-scale overdense regions. Compared to underdense regions, these areas contain more matter and initially more neutral hydrogen, which consequently glow more brightly in the 21 cm emission. This leads to a positive 21cm-NIRB cross-correlation on large scales, as shown by the green-thin lines in Figure 5. As reionization proceeds, the galaxies quickly ionize their overdense surroundings completely. These overdense regions hence dim in 21 cm emission. On the other hand, the large-scale underdense regions are still mostly free of galaxies and roughly maintain their initial 21 cm brightness temperature. Therefore, an anti-correlation is demonstrated during the middle and late stages of reionization, as shown by the red-thick lines in Figure 5. Overall, the amplitude of cross-correlation grows even faster with redshift after the reionization is half completed, and the maximum is reached when $z=7$ with $\bar{x}_{\rm{i}}\approx0.97$. This marks a rise of the cosmic structure formation which is responsible for the reionization of neutral hydrogen at late times.

In lower panels, we show the cross-correlation coefficient for various 21 cm redshift slices. One can see that the absolute value of $r_{\rm{corr}}$ reaches a maximum of $\sim 0.05$ when $\bar{x}_{\rm{i}}\sim 0.6$ for the two NIR bands considered here. Hence we infer that only a small part of the NIRB cross-correlates with the 21 cm emission from the EoR. The result is easy to understand: the NIRB contains remnant light of high-redshift galaxies. Specifically, the redshift intervals of the NIRB integral are $6\leqslant z \leqslant17$ and $6\leqslant z \leqslant30$ for H-band and L-band, respectively. However, the bandwidth of 21 cm signal is significantly narrow $\Delta\nu=0.1\,$MHz, corresponding to $\Delta z=7\times10^{-3}$ at $z=9$. In cross-correlation analysis, most of the NIR light comes from redshifts different from that of the measured 21 cm background. Fortunately, a cross-correlation does exist between these two radiation backgrounds. Any detection of this cross signal in future experiments could tell us properties of early stars and physics of reionization process.


    \begin{deluxetable*}{p{5cm}cccc}

    \tablewidth{40pc}
    \tablecaption{Parameters that we adopt for LOFAR, SKA and CIBER}

    \tablehead{ & \colhead{LOFAR} & \colhead{SKA} & \colhead{CIBER-I} & \colhead{CIBER-II}}
    \startdata

    Pixel size, $\theta_b$ & 5' & 1' & 10'' & 1'' \\
    System temperature, $T_{\rm sys}$ & 500 K & 500 K &  &  \\
    Sky coverage, $f_{\rm sky}$ & 0.1 & 0.5 & $4\times10^{-4}$ & 0.1 \\
    Number of antennae, $N_{\rm ant}$ & 50 & 200 &  &  \\
    Spectral resolution, $\Delta \nu$ & 0.1 MHz & 0.1 MHz &  & \\
    Integration time, $t$ & 1000 hr & 1000 hr & 1 hr & 1 hr \\
    Pixel sensitivity, $\sigma_{\rm pix}$ &  &  & $10\,\rm {nW m^{-2} sr^{-1}}$ & $1\,\rm {nW m^{-2} sr^{-1}}$ \\

     \enddata

    \tablecomments{For NIR observations, we assume the integration time $t=1$ hr for a optimistic prediction. Moreover, we assume $f_{\rm sky}=0.1$ for CIBER-II survey which is a hypothetical extension of the current CIBER. The other instrumental parameter values come from \citet{Mellema12} for LOFAR and SKA, and \citet{Cooray09} and \citet{Zemcov11} for CIBER-I and CIBER-II.}
    \end{deluxetable*}

\section{Detectability}

Here we consider measurement errors on the expected cross-correlation signal. To get a sense of the precision which we are able to achieve, we compute analytic approximations of measurement uncertainties in a simple way that is often accurate enough for theoretical studies \citep{Knox95}. First, we show errors on detections of the auto-correlation power spectra of 21 cm and NIR backgrounds. Then we estimate the associated errors on the determination of their cross-correlation. We discuss the following experiments in this paper: LOFAR, SKA, CIBER-I \footnote{Cosmic Infrared Background Experiment, http://physics.ucsd.edu/~bkeating/CIBER.html} and CIBER-II. Table 2 lists the instrumental parameter values used for our sensitivity calculation. Unless otherwise stated, the parameters we adopt come from \citet{Mellema12} for LOFAR and SKA, and \citet{Cooray09} and \citet{Zemcov11} for CIBER-I and CIBER-II.

Neglecting systematic effects such as foreground subtraction, errors on a measurement of the 21 cm angular power spectrum $C^{\psi\psi}_{\ell}$ at a particular mode $\ell$ can be given by \citep{Knox95}
\begin{eqnarray}
\Delta C^{\psi\psi}_{\ell}=\sqrt{\frac{2}{(2\ell+1)f_{\rm sky}}}\bigg[C^{\psi\psi}_{\ell}+N^{\psi\psi}_{\ell}\bigg],
\end{eqnarray}
with the instrumental noise
\begin{eqnarray}
N^{\psi\psi}_{\ell}=\omega^{-1}\,e^{\theta^2_b \ell(\ell+1)}=\sigma^2_{\rm pix}\,\Omega_{\rm pix}\,e^{\theta^2_b \ell(\ell+1)}.
\end{eqnarray}
Here we take the pixel solid angle to be $\Omega_{\rm pix}=\theta_{\rm fwhm} \times \theta_{\rm fwhm}$. Moreover, $f_{\rm{sky}}$ is the observed fraction of the sky, $\theta_b$ is the standard deviation for a Gaussian beam function in the experiment, and $\sigma_{\rm pix}$ is defined as the RMS noise in each pixel \citep{Mao08}. The above approximation is formally correct if the 21 cm temperature field is a statistically Gaussian distribution. Unfortunately, this condition does not strictly hold. However, we ignore these and use this relationship here since it allows considerable simplification.

On the NIRB side, the analogous errors can be written as \citep{Cooray04b}
\begin{eqnarray}
\Delta C^{\phi\phi}_{\ell}=\sqrt{\frac{2}{(2\ell+1)f_{\rm sky}}}\bigg[C^{\phi\phi}_{\ell}+N^{\phi\phi}_{\ell}\bigg],
\end{eqnarray}
where $N^{\phi\phi}_{\ell}=\sigma^2_{\rm pix}\,\Omega_{\rm pix}\,e^{\theta^2_b \ell(\ell+1)}$ accounts for the instrumental noise associated with the NIRB measurements. The errors in measurements of the 21 cm and NIRB power spectra with LOFAR and CIBER-I are displayed in the top and bottom panels of Figure 3, respectively. While it is promising for LOFAR to detect the angular power spectrum of 21 cm fluctuations after an integration time of 1000 hours, a significant detection of the NIRB angular power spectrum turns to be difficult due to the limited field of view of CIBER-I.

\begin{figure*}
\begin{center}
\includegraphics[angle=270, scale=0.7]{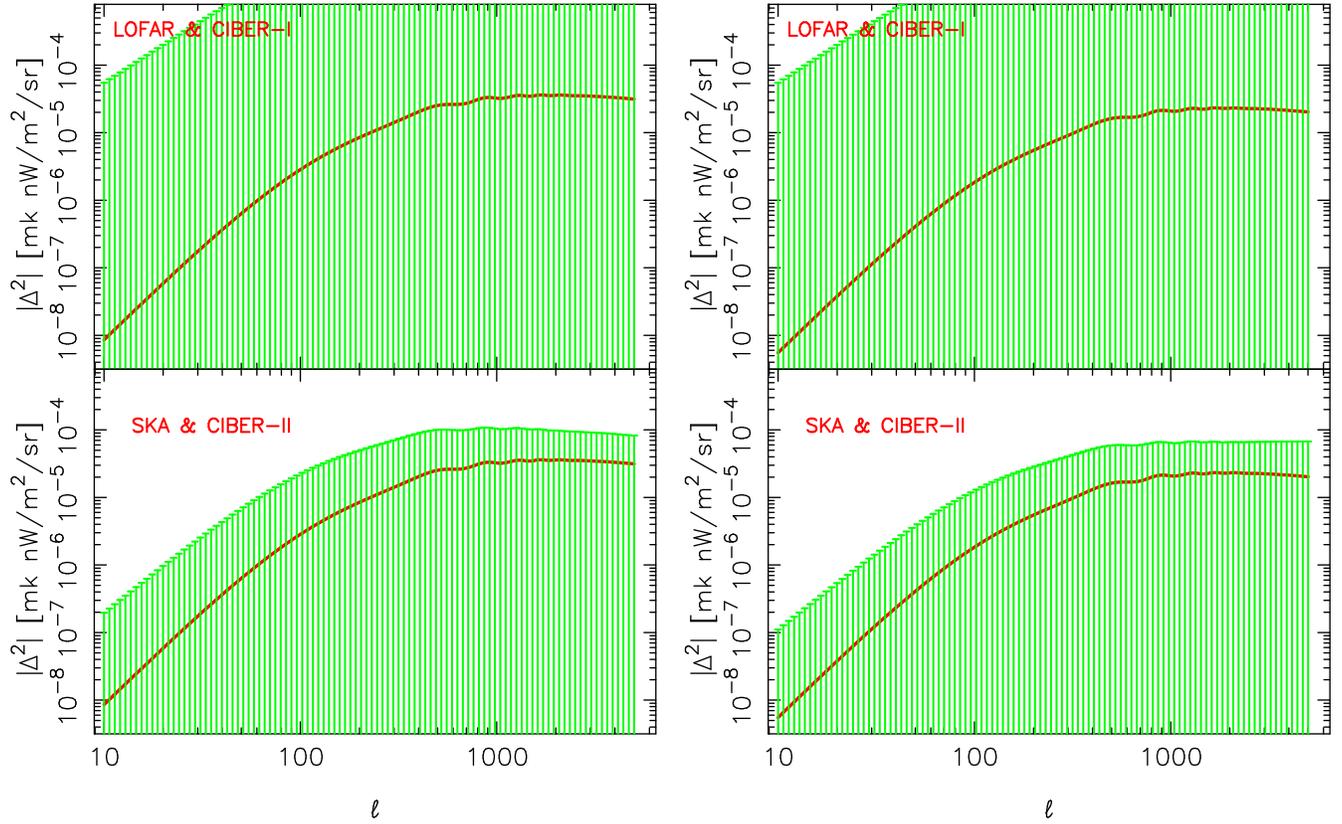}
\caption{21 cm-NIRB angular power spectrum and measurement variance. The 21 cm signal at redshift $z=9$ is correlated against the high-redshift component of the NIRB observed at H-band (left) and L-band (right) respectively. We consider two different schemes for combined observations: LOFAR \& CIBER-I (top) as well as SKA \& CIBER-II (bottom). The instrumental parameters adopted in our sensitivity calculation are listed in Table 2.}
\end{center}
\end{figure*}

Finally we can estimate the associated errors in cross-correlation measurements through \citep{Song03}
\begin{eqnarray}
\Delta C^{\psi\phi}_{\ell}&=&\sqrt{\frac{2}{(2\ell+1)f_{\rm sky}}}\Bigg[\Big(C^{\psi\phi}_{\ell}\Big)^2 \nonumber \\
&&+\Big(C^{\psi\psi}_{\ell}+N^{\psi\psi}_{\ell}\Big)\,\Big(C^{\phi\phi}_{\ell}+N^{\phi\phi}_{\ell}\Big)\Bigg]^{1/2}.
\end{eqnarray}
Note that here $f_{\rm sky}$ is the fraction of sky observed by both experiments for cross-correlation studies. In Figure 6, we show the expected cross-correlation signal and its measurement variance. The 21 cm radiation background is measured at $z=9$ with a frequency bin $\Delta \nu=0.1$ MHz. We correlate this 21 cm signal against the entire NIRB from the EoR at H-band (left) and L-band (right) observations, respectively. The top panels make it clear that we are unable to extract meaningful information from cross-correlation detections with current telescopes, like LOFAR and CIBER-I. The errors expected from surveys of a larger covering area seem to be smaller especially at small scales (bottom panels). It indicates that an experiment with poorer sky coverage will be less precise at every value of $\ell$, since the limited number of available modes will lead to large sample variance. For this reason, it is suggested that a cross-correlation experiment should make the covered sky area as large as possible in a fixed observing time in order to reduce the sample variance which dominates measurement errors on large scales. In addition, we find that the measurement errors still exceed the expected cross-correlation signal at $\ell < 5000$ even if $f_{\rm{sky}}=0.1$.

As we see in Figure 7, the measurement errors could be reduced significantly by increasing the bandwidth (i.e. frequency resolution) in 21 cm experiment. Here we show results with $\Delta\nu=10\,$MHz, which is 100 times larger than that in Figure 6. Comparing Figure 7 to Figure 6, there is no significant difference in the cross-correlation signal $C^{\psi\phi}_{\ell}$, while the overall level of measurement errors has been reduced obviously. Especially for the combined observation with SKA and CIBER-II, the reduction in $\Delta C^{\psi\phi}_{\ell}$ is dramatic, greatly expanding the available range of scales. This is largely due to two reasons: 1) the 21 cm power spectrum $C^{\psi\psi}_{\ell}$ decreases with the amount of $\Delta\nu$, 2) the RMS noise $\sigma_{\rm{pix}}$ in 21 cm measurements is proportional to the factor $1/\sqrt{\Delta\nu}$, and thus increasing the bandwidth $\Delta\nu$ decreases $N^{\psi\psi}_{\ell}$. From Equation (34), one can see that lower levels of $C^{\psi\psi}_{\ell}$ and $N^{\psi\psi}_{\ell}$ will make smaller errors $\Delta C^{\psi\phi}_{\ell}$ in cross-correlation detections.

\section{Discussion}

Cross-correlation analysis could provide an important complement to the existing reionization studies, yielding new insights into the topology of reionization and eliminating some of the difficulties related to foreground removal. In this paper, we investigated the cross-correlation between the cosmic 21 cm and NIR radiation backgrounds, and further quantified the likelihood of measuring this cross power spectrum with future EoR experiments.

We adopted a well-studied halo approach to describe the matter distribution during reionization. For first galaxies, the halo occupation distribution model was used to examine how these galaxies occupy dark matter halos. Just as shown in Figure 2, we considered two stellar populations in the star formation model: metal-free (Pop III) stars and metal-poor (Pop II) stars. Since the UV radiation from high-redshift galaxies is expected to be present in the NIR background light, we estimated the NIRB intensity for five different emission processes including stellar emission, Lyman-$\alpha$ line, and free-free, free-bound and two photon emission. Collecting all emission from these processes, we obtained the NIR mean emissivity of the stellar nebulae and neglected the same of the IGM. The reionization process was described through the hydrogen ionization fraction as a function of redshift. To test if the reionization history is consistent with the WMAP data, we further estimated the optical depth to electron scattering and got $\tau=0.084$.

\begin{figure*}
\begin{center}
\includegraphics[angle=270, scale=0.7]{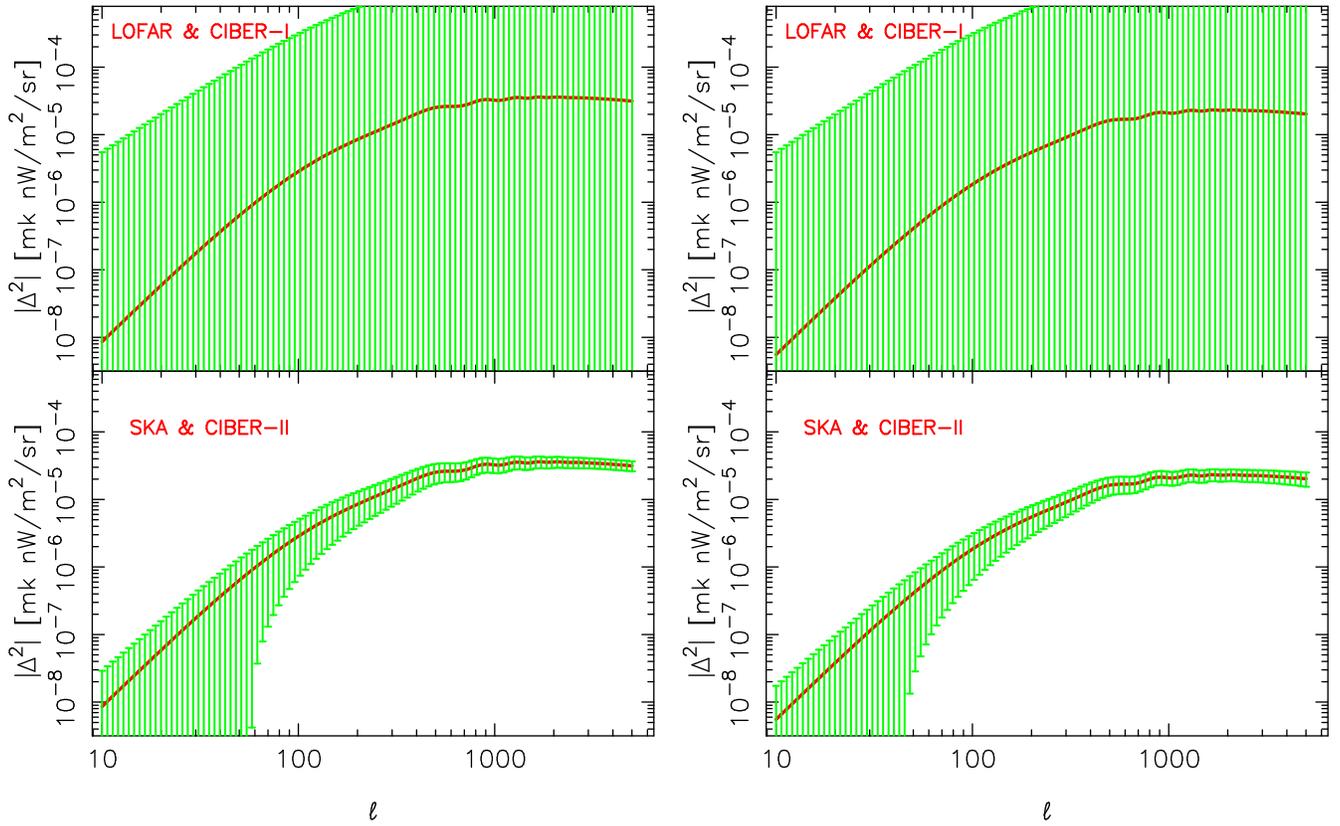}
\caption{Same as Figure 6, but with larger bandwidth in 21 cm experiments $\Delta \nu=10$MHz.}
\end{center}
\end{figure*}

Based on the constructed reionization model, we explored the cross-correlation signal between the cosmic 21 cm and NIR radiation backgrounds during the EoR. Since the appropriate theoretical model used to calculate $P_{\rm{x,gal}}$ may be not valid on small scales, we will treat more of the physics properly through numerical simulations in the near future. Here we therefore focus on the large-scale behavior of the 21 cm-NIRB cross power spectrum. In our fiducial model, the 21 cm background is measured at an individual redshift with a frequency bin $\Delta\nu=0.1 \rm{MHz}$, and we cumulate the NIR background light from the entire epoch of reionization at the observed H-band and L-band. We found that the 21 cm emission background is positively correlated with the cumulative NIRB in the early phases of reionization. However, by $z\sim11$, at which point the mean ionization fraction of neutral hydrogen is $\sim 0.3$, these two radiation backgrounds become anti-correlated on large scales. Furthermore, we study the redshift and scale dependence of the cross-correlation signal. We showed that the amplitude of the cross power spectrum changes obviously as reionization proceeds, while a similar shape is preserved. Specifically, the overall level of the cross-correlation signal reaches the maximum when the cumulative NIRB is correlated with the 21 cm background at redshift $z=7$ with $\bar{x}_{i}\approx0.97$. The absolute value of the cross power spectrum $|\Delta_{\rm{21,NIR}}^2|$ increases by about 4 orders of magnitude over the scale range $10\leqslant \ell \leqslant 5000$. Note that the represented cross power spectrum shows no evidence of a turnover that was claimed to exist in \citet{Slosar07}. Different from cross power spectrum, the cross-correlation coefficient $r_{\rm{corr}}$ changes slightly with $\ell$. On the other hand, the $r_{\rm{corr}}$ grows obviously with redshift during the middle and late phases of reionization. The maximum of $r_{\rm{corr}}$ appears when $\bar{x}_{i}\sim0.6$, which is only $r_{\rm{corr}}\sim0.05$ . To a certain extent, this is induced by the narrow bandwidth of 21 cm measurements $\Delta \nu=0.1$ MHz. In the discussed scenario, most of the NIRB radiation is uncorrelated with the redshifted 21 cm line emission and acts as a source of noise. Different from our results, \citet{Fernandez14} found a quite strong cross-correlation between these two radiation backgrounds. This is mainly because the redshift intervals of their 21 cm maps are much larger than ours. Their simulations showed a positive correlation at very early times and an anti-correlation at the mid and late reionization times. The anti-correlation is the strongest when the Universe is $50\%$ ionized or more. They drew some qualitative conclusions agree with the results obtained in our work.

In addition, we computed an analytic approximation of measurement uncertainties for several upcoming experiments. The accuracy with which the cross power spectrum can be measured directly depends on the sampling variance. The greatest problem comes from the NIR observation: the early experiments like CIBER-I are limited by a very small fraction of the sky being observed. As shown in Figure 7, a significant detection of the 21 cm-NIRB cross power spectrum may be achieved by combing SKA and CIBER-II surveys, provided that the integration time independently adds up to 1000 and 1 hours for 21 cm and NIR observations, and that the sky coverage fraction of CIBER survey is extended from $4\times10^{-4}$ to 0.1. Surveys required for cross-correlation measurements are clearly very challenging for current telescopes, but rapid progress is being made in this direction as deep, widefield surveys are being designed to study the high-redshift universe.

Since the 21 cm and NIR radiation backgrounds are non-Gaussian, the auto power spectra alone provide an incomplete description of the fields' statistical properties. Detection of cross-correlation is therefore quite valuable, which would not only help confirm the exact intensity of the ``missing'' NIR background from high-redshift sources but also offer additional information about the reionization process. Developing the theoretical analysis of cross-correlation is particularly timely, since it can be used to forecast important constraints and eventually interpret the results of future observations.

\acknowledgments{We thank the anonymous referee for valuable comments and suggestions. We also thank Xiang-Ping Wu, Yan Gong and Asantha Cooray for helpful discussions. This work was supported by the National Science Foundation of China, under Grant No. 11003019. This work was supported in part by the 973 Program grant No. 2013CB837900, NSFC grant No. 11261140641, and CAS grant No. KJZD-EW-T01.}

\end{document}